\documentclass[twocolumn]{aastex631}

\newcommand*{\dprime}{^{\prime\prime}\mkern-1.2mu}

\usepackage{graphicx}
\usepackage{subfigure}

\definecolor{darkorange}{rgb}{1.0, 0.55, 0.0}

\definecolor{bluegray}{rgb}{0.4, 0.6, 0.8}

\definecolor{magenta}{rgb}{1.0, 0.0, 1.0}

\definecolor{indigo(web)}{rgb}{0.29, 0.0, 0.51}

\usepackage{soul}
\DeclareUnicodeCharacter{0308}{`}
\DeclareUnicodeCharacter{0301}{``}

\begin{document}

\title{\emph{JWST} MIRI MRS Images Disk Winds, Water, and CO in an Edge-On Protoplanetary Disk}

\author[0000-0003-2631-5265]{Nicole Arulanantham}
\affil{Space Telescope Science Institute, 3700 San Martin Drive, Baltimore, MD 21218, USA}

\author[0000-0003-1878-327X]{M.K. McClure}
\affil{Leiden Observatory, Leiden University, PO Box 9513, NL–2300 RA Leiden, The Netherlands}

\author[0000-0001-7552-1562]{Klaus Pontoppidan}
\affil{Jet Propulsion Laboratory, California Institute of Technology, 4800 Oak Grove Drive, Pasadena, CA 91109, USA}

\author[0000-0002-6881-0574]{Tracy L. Beck}
\affil{Space Telescope Science Institute, 3700 San Martin Drive, Baltimore, MD 21218, USA}

\author[0000-0002-0377-1316]{J.A. Sturm}
\affil{Leiden Observatory, Leiden University, PO Box 9513, NL–2300 RA Leiden, The Netherlands}

\author[0000-0001-6307-4195]{D. Harsono}
\affil{Institute of Astronomy, Department of Physics, National Tsing Hua University, Hsinchu, Taiwan}

\author[0000-0001-9344-0096]{A.C.A. Boogert}
\affil{Institute for Astronomy, University of Hawai’i at Manoa, 2680 Woodlawn Drive, Honolulu, HI 96822, USA}

\author[0000-0001-8233-2436]{M. Cordiner}
\affil{NASA Goddard Space Flight Center}

\author[0000-0003-1197-7143]{E. Dartois}
\affil{Institut des Sciences Mol\'{e}culaires d’Orsay, CNRS, Univ. Paris-Saclay, 91405 Orsay, France}

\author[0000-0001-7479-4948]{M.N. Drozdovskaya}
\affil{Physikalisch-Meteorologisches Observatorium Davos und Weltstrahlungszentrum (PMOD/WRC), Dorfstrasse 33, CH-7260, Davos Dorf, Switzerland}

\author[0000-0001-9227-5949]{C. Espaillat}
\affil{Institute for Astrophysical Research, Department of Astronomy, Boston University, 725 Commonwealth Avenue, Boston, MA 02215, USA}

\author{G.J. Melnick}
\affil{Center for Astrophysics | Harvard \& Smithsonian, 60 Garden St., Cambridge, MA 02138, USA}

\author[0000-0003-4985-8254]{J.A. Noble}
\affil{Physique des Interactions Ioniques et Mol\'{e}culaires, CNRS, Aix Marseille Univ., 13397 Marseille, France}

\author[0000-0002-9122-491X]{M.E. Palumbo}
\affil{INAF - Osservatorio Astrofisico di Catania, via Santa Sofia 78, 95123 Catania, Italy}

\author[0000-0001-8102-2903]{Y.J. Pendleton}
\affil{Department of Physics, University of Central Florida, Orlando, FL 32816, USA}

\author[0000-0002-7914-6779]{H. Terada}
\affil{National Astronomical Observatory of Japan (NAOJ)}

\author[0000-0001-7591-1907]{E.F. van Dishoeck}
\affil{Leiden Observatory, Leiden University, PO Box 9513, NL–2300 RA Leiden, The Netherlands}

\begin{abstract}

We present JWST MIRI MRS observations of the edge-on protoplanetary disk around the young sub-solar mass star Tau~042021, acquired as part of the Cycle 1 GO program “Mapping Inclined Disk Astrochemical Signatures (MIDAS).” These data resolve the mid-IR spatial distributions of H$_2$, revealing X-shaped emission extending to $\sim$200 au above the disk midplane with a semi-opening angle of $35 \pm 5^{\circ}$. We do not velocity-resolve the gas in the spectral images, but the measured semi-opening angle of the H$_2$ is consistent with an MHD wind origin. A collimated, bipolar jet is seen in forbidden emission lines from [Ne II], [Ne III], [Ni II], [Fe II], [Ar II], and [S III]. Extended H$_2$O and CO emission lines are also detected, reaching diameters between $\sim 90$ and 190 au, respectively. Hot molecular emission is not expected at such radii, and we interpret its extended spatial distribution as scattering of inner disk molecular emission by dust grains in the outer disk surface. H I recombination lines, characteristic of inner disk accretion shocks, are similarly extended, and are likely also scattered light from the innermost star-disk interface. Finally, we detect extended PAH emission at 11.3$\mu$m co-spatial with the scattered light continuum, making this the first low-mass T Tauri star around which extended PAHs have been confirmed, to our knowledge. MIRI MRS line images of edge-on disks provide an unprecedented window into the outflow, accretion, and scattering processes within protoplanetary disks, allowing us to constrain the disk lifetimes and accretion and mass loss mechanisms.

\end{abstract}


\section{Introduction}

Planets form within circumstellar disks of gas, dust, and ice, where they must accumulate material before it is removed via outflow mechanisms and accretion onto the central stars. Observations suggest that this process takes at most a few Myr, as determined by the age at which the sub-mm dust and gas content in disks decreases significantly (see e.g., \citealt{Pascucci2016, Barenfeld16, Simon19}). Models of viscously evolving disks predict a corresponding decline in mass accretion rates, as the gas reservoir is also depleted over time \citep{LyndenBell74, Hartmann98}. However, this picture is complicated by significant scatter in the observed relationship between mass accretion rates $\left( \dot{M}_{\rm{acc}} \right)$ and disk dust masses $\left(M_{\rm{disk}} \right)$ \citep{Hartmann98, Manara23, Betti23}, which indicates that additional physical mechanisms are required to drive mass loss and set the maximum lifetimes of protoplanetary disks \citep{pascucci2023}.    

Recent work demonstrates that the relationship between $\dot{M}_{\rm{acc}}$ and $M_{\rm{disk}}$, including the scatter, can be reproduced by models where magnetohydrodynamic (MHD) winds are the primary drivers of angular momentum transfer that enables ongoing accretion \citep{tabone2022b, somigliana23}. In this framework, a disk population begins with a distribution of initial disk sizes and viscosities \citep{tabone2021}. A magnetized disk wind, removing mass at approximately the same rate as accretion, can then drive disk dispersal over a timescale consistent with observations \citep{Blandford1982, Cabrit1999, Bai2013, Lesur21}.

Although MHD wind models can successfully reproduce the decline in $\dot{M}_{\rm{acc}}$ and $M_{\rm{disk}}$ over time, unambiguous distinction between outflow-launching mechanisms in the inner disk remains elusive (jets, MHD and photoevaporative winds; see e.g., \citealt{pascucci2023} for a recent review). Outflow signatures are readily identified in high-resolution spectroscopy, which reveals blue-shifted emission and absorption lines from atomic transitions, including [O I], [Ne II], [N II], [S II], He I, and C II \citep{Edwards87, Kwan1995, Pascucci2009, Banzatti2019, Haffert20, Xu21, Fang18, Fang23_US, CampbellWhite23}. The central velocities and FWHM of the line profiles, which can be interpreted as average outflow velocities and radial locations within a Keplerian disk, are typically used to distinguish between an origin in photoevaporative and/or MHD winds. For example, a more compact emitting region ($r \sim 0.1-0.2$ au) is assumed to be characteristic of an MHD wind \citep{CampbellWhite23}. Particularly the line flux ratios of [Ne II] 12.814 $\mu$m and [O I] 6300 \AA \, in low-velocity photoevaporative winds increase as emission from warm, inner disk dust declines \citep{Pascucci2020}. A similar correlation is detected between the inner disk dust content and the morphology of accretion-generated Ly$\alpha$ \citep{Arulanantham2023} and H$\alpha$ emission lines \citep{Fang23_US}, indicating that the relative contributions of MHD and photoevaporative winds and accretion to the total mass loss rate may evolve over time.

Searches for spatially extended gas from protoplanetary disk winds have recently detected [O I] and Ly$\alpha$ emission from the face-on system TW Hya \citep{Fang2023, Chang23}. The [O I] morphology is reproduced by models of a radially confined MHD wind launched from within the dust sublimation zone, while the Ly$\alpha$ distribution is consistent with resonant scattering from the accretion shocks through H I within both a wind and the disk itself. Spectroastrometry of the same tracer in RU Lupi and AS 205 N \citep{Whelan2021} reveals a more extended emitting region consistent with the superposition of a wide-angled MHD disk wind ($0 < |v| < 40$ km s$^{-1}$) and an accelerating jet ($90 < |v| < 230$ km s$^{-1}$). Spatially extended near-infrared emission from ro-vibrational H$_2$ lines has also been identified from T Tauri systems \citep{Eisloffel2000, Duchene05, Beck2008, Beck2019}, as expected from high-resolution 1-D spectroscopy that showed blue-shifted emission line profiles \citep{Beckwith78, Bary2003, Gangi2020}. The emission lines originate in gas with temperatures between $T \sim 1800-2300$ K and yield line flux ratios consistent with shock excitation \citep{Eisloffel2000, Beck2008}. Mid-infrared emission from rotational quadrupole transitions of H$_2$ has been detected at temperatures between $T \sim 100-200$ K, showing line profiles consistent with extended emission \citep{Thi2001, Bitner08}. Still, it remains observationally challenging to distinguish whether MHD or photoevaporative winds are responsible for driving the detected outflows, as the wind launching radii and velocity structure remain spatially and spectrally unresolved at the distance to nearby star-forming regions (see e.g., \citealt{Beck2008, Coleman24}). Clues may come from spatially resolved observations of [Ne II], where the emission line luminosity increases as dust cavities are cleared in the disk and the contribution from photoevaporative winds becomes more prominent \citep{Pascucci2020}.

In this paper, we report the detection of spatially extended rotational H$_2$, H$_2$O, and CO emission from an edge-on protoplanetary disk with \emph{JWST}-MIRI/MRS, using Integral Field Unit (IFU) observations. We describe the target, observations, and our method for constructing line and continuum images in Section 2. In Section 3, we compare the morphology of the molecular lines to emission lines from forbidden atomic transitions and H I Humphreys $\alpha$ emission, which allows us to consider winds, jets, and accretion as drivers of mass loss within the same system. We also compare the line images to the spatial distribution of mid-infrared scattered light to constrain the origin of the extended molecular gas emission. The kinematics of the extended emission are discussed in Section 4, and we conclude in Section 5. 

\section{Target and Observations} 
\label{targets}
\subsection{Description of Observing Program}
The target object, 2MASS J04202144+2813491 (hereinafter Tau~042021; RA 04:20:21.44, Dec +28:13:49.17) surrounds a low-mass star \citep[M1, 0.27\, $M_{\odot}$; ][]{Simon19} star at a distance of $\sim$140 pc surrounded by a 5$\dprime$ edge-on disk ($>$85\degr) first seen at optical wavelengths with the Canada-France-Hawaii Telescope (CFHT; \citealt{luhman2009}) and later imaged with the Advanced Camera for Surveys (ACS) onboard the \emph{Hubble Space Telescope} \citep{Stapelfeldt2014, Villenave2020}. It is also one of the few edge-on disks for which both the radial and vertical structure have been resolved in ALMA observations \citep[Band 4, 3.67$\dprime$ by 0.2$\dprime$,][]{Villenave2020, Villenave2023}. New \emph{JWST} images of the system acquired with NIRCam and MIRI reveal a well-mixed population of 10 $\mu$m dust grains, which are coupled to the gas in surface layers down to the midplane; the same images also show an X-shape in the F770W and F1280W filters, which may be associated with a disk wind \citep{Duchene24}. A bipolar, collimated jet is detected as well, using the F606W filter on \emph{HST}-ACS \citep{Duchene14} and \emph{JWST}-MIRI imaging at 12.8 $\mu$m \citep{Duchene24}. We selected Tau~042021 for the MIDAS program on the basis of its 6 and 6.8 $\mu$m ice features, along with strong H$_2$ and [Ne II] 12.8 $\mu$m gas emission lines and broad 11.3 $\mu$m PAH emission that were visible in its {\it Spitzer} IRS spectrum (PI: J.R. Houck; PID: 50053; see e.g., \citealt{furlan2011, PaperII, Duchene24}). 

Our JWST MIRI MRS observations were obtained on February 13th, 2023 as part of PID 1751 (PI: M. McClure). Since the target is extended and has HST coordinates, we did not use target acquisition. Although the disk diameter is 5$\dprime$ at optical wavelengths, which is larger than the MIRI detector between 4.9-7.65 $\mu$m, we modeled its spatial extent using radiative transfer models fit to the HST image and {\it Spitzer} spectrum and determined that the mid-infrared scattered light image would fit within MIRI's field-of-view (FOV). We used a 4-point dither pattern optimized for an extended source, with each dither exposure consisting of one integration of 69 groups, read using the FASTR1 readout pattern, for total on-source exposure times of 6594 s in each of the MIRI channels. This integration time was selected to get a S/N of 10 per spaxel on the central 3 $\times$ 5 spaxel region of our simulated disk model, which would enable detection of CH$_4$ and other weak ices after rebinning. Since the source is extended and large, a dedicated sky background was obtained with dithers of the same number of groups, but a 2-point dither pattern to reduce the non-science time. The final 3D IFU cube was produced using version 11.17.1 of the Calibration References Data System (CRDS) file selection software in the \emph{JWST} Science Calibration Pipeline (CRDS context ``jwst\_1174.pmap"), with Stage 3 processing utilized to combine data from all dither positions \citep{bushouse23}. 1D spectra were extracted from the eastern and western lobes of the disk and calibrated using the JDISCS extractor \citep{Pontoppidan2023}.

The spectral resolution of the MRS changes as a function of wavelength. While in-flight observations during the instrument commissioning were used to derive resolving powers between $R \sim 1500-3500$ \citep{Jones2023}, calibration against high-resolution ground-based spectroscopy of the protoplanetary disk system FZ Tau are consistent with $R \sim 2000-3000$ \citep{Pontoppidan2023}. Using the coefficients for each sub-band provided in \citealt{Pontoppidan2023}, we expect $\Delta v \sim 106$ km s$^{-1}$ at 5 $\mu$m and $\Delta v \sim 123$ km s$^{-1}$ at 28 $\mu$m. 

Observations are acquired in four different channels: 4.9-7.65 $\mu$m (Channel 1), 7.51-11.7 $\mu$m (Channel 2), 11.55-17.98 $\mu$m (Channel 3), and 17.7-27.9 $\mu$m (Channel 4). The smallest FOV and highest spatial resolution are available at shorter wavelengths in Channel 1 (3.2''x3.7'', 0.196'' pixel$^{-1}$). The FOV increases to 6.6''x7.7'' in Channel 4, and the spatial resolution decreases to 0.273'' pixel$^{-1}$. Characterization of the MIRI PSF for extended sources in the along-slice and across-slice directions is currently ongoing and will be required to compare spatially resolved emission between channels with different pixel scales. For point sources, the average PSF FWHMs increase with wavelength, such that $\theta = 0.033 \left(\lambda \, \mu \rm{m} \right) + 0.106''$ \citep{law2023, argyriou23}.  

\subsection{Construction of Line and Continuum Images}
To explore the spatial extents of the gas emission lines, we produced continuum-subtracted spectral images for the detected transitions. The 2-D local dust continuum was measured as the median surface brightness per spaxel in frames selected from emission line-free wavelengths along the spectral axis (see e.g., \citet{Pontoppidan2023, worthen2024}). Approximately 10 frames were used to estimate the continuum, spanning $\pm 0.007$ $\mu$m on either side of the emission line. This background image was subtracted from each of the 4-5 frames containing the line emission, which were then stacked to produce the final image. We note that the target was not centered on the detector during the observations, requiring an additional correction to shift the disk to the zero position in the line images presented here. Since the MIRI-MRS IFU cubes produced by the \emph{JWST} calibration pipeline are registered to the World Coordinate System, we determined the correction as the difference between the RA and Dec of Tau~042021 and the image center (1-3 pixels, depending on the channel). We do not resample the images, so any remaining deviations from the zero point are likely due to the large MIRI-MRS spaxel sizes.

For both the CO and H$_2$O line series, all detected features were co-added to produce the final line surface brightness maps. The heights vertical to the disk and diameters along the midplane of each emitting region were measured as the area within which the surface brightness is greater than $40\% \pm 2.5\%$ of the peak surface brightness in the spectral images (see Table \ref{tab:line_IDs}). The 40\% threshold was chosen instead of a FWHM measurement, as this is the limit where emission from both sides of the disk can be distinguished from the residual background in the line images (see e.g., \citealt{Beck2008}). The uncertainties are derived from the surface brightness limit where the emitting region shifts into the next adjacent spaxels. We discuss the observed gas-phase morphologies in the following sections. 

\section{Results} 

\begin{figure*}
\centering
\includegraphics[width=\linewidth]{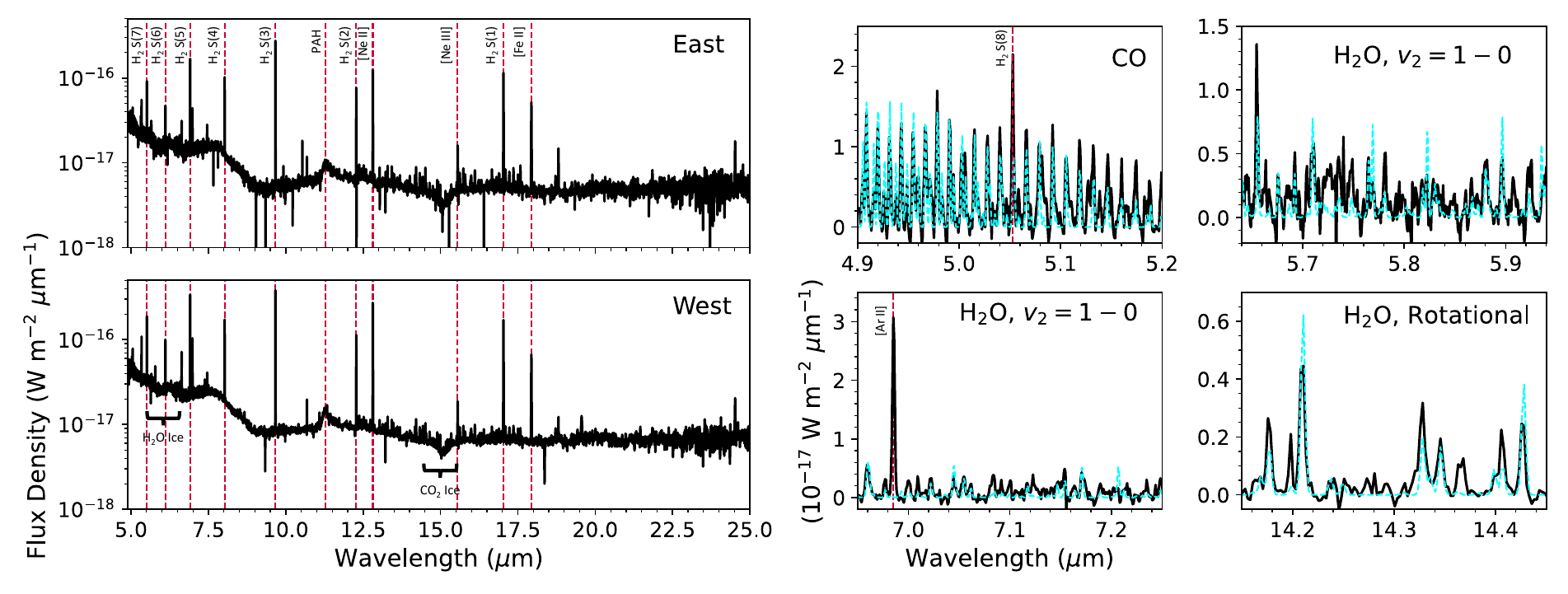}
\caption{\emph{Left:} 1-D MIRI MRS spectrum of Tau~042021, extracted from the eastern (top) and western (bottom) reflection nebulae. We detect H$_2$ emission lines and forbidden emission lines from [Ne II], [Ne III], and [Fe II] on both sides of the disk. H$_2$O and CO$_2$ ice absorptions are also identified, along with PAH emission near 11 $\mu$m. \emph{Right:} Zoomed in spectra of ro-vibrational CO, ro-vibrational H$_2$O, and rotational H$_2$O emission lines from the eastern reflection nebula, with LTE slab models overlaid in cyan \citep{salyk2022}. The CO model has a temperature and column density of $T = 1500$ K, $\log N = 17.5$ cm$^{-2}$, and the rotational H$_2$O emission lines are consistent with $T = 850$ K, $\log N = 15.0$. The ro-vibrational lines are shown with the rotational model scaled down by a factor of 20 in flux.}
\label{fig:1Dspec}
\end{figure*}

\begin{figure*}
\centering
\includegraphics[width=\linewidth]{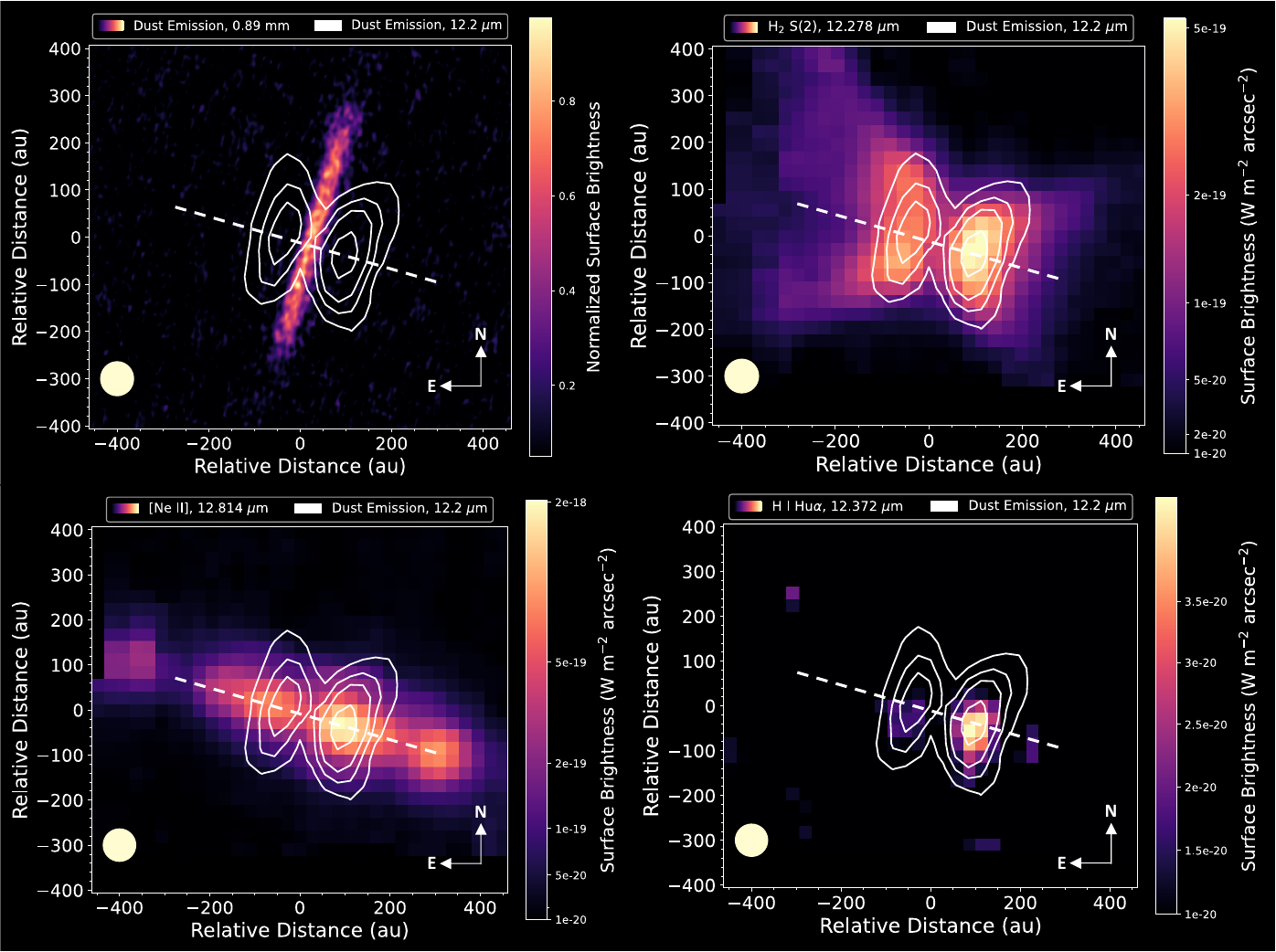}
\caption{MIRI MRS images of Tau~042021, showing the ALMA 0.89 mm dust continuum (top left; \citealt{Villenave2020}) and emission lines from H$_2$ S(2) 12.278 $\mu$m (top right), [Ne II] 12.814 $\mu$m (lower left), and H I Humphreys $\alpha$ 12.372 $\mu$m (bottom right) detected with the MIRI MRS. White contours in each image show the mid-IR scattered light continuum at 12.2 $\mu$m, and the white dashed lines show the position angle measured from the ALMA continuum image ($\rm{PA} = -16^{\circ}$; \citealt{Villenave2020})}. Images have been scaled using an asinh stretch, to maximize the contrast between the background noise and the high S/N emission from the source. Beige circles in the bottom left corner of each panel represent the average theoretical FWHM of the PSF, calculated from the relationship reported in \citet{law2023}. The observed spatial distributions are consistent with $\mu$m-size grains in the eastern and western reflection nebulae (12 $\mu$m continuum), a disk wind (H$_2$ S(2)), bipolar jet ([Ne II]) and scattered light from accretion shocks or the base of the jet (H I). We note that the position angle of the jet is tilted by $\sim2.5^{\circ}$ relative to the disk.
\label{fig:fig1_dustH2NeIIHI}
\end{figure*}

\begin{figure*}
\centering
\includegraphics[width=\linewidth]{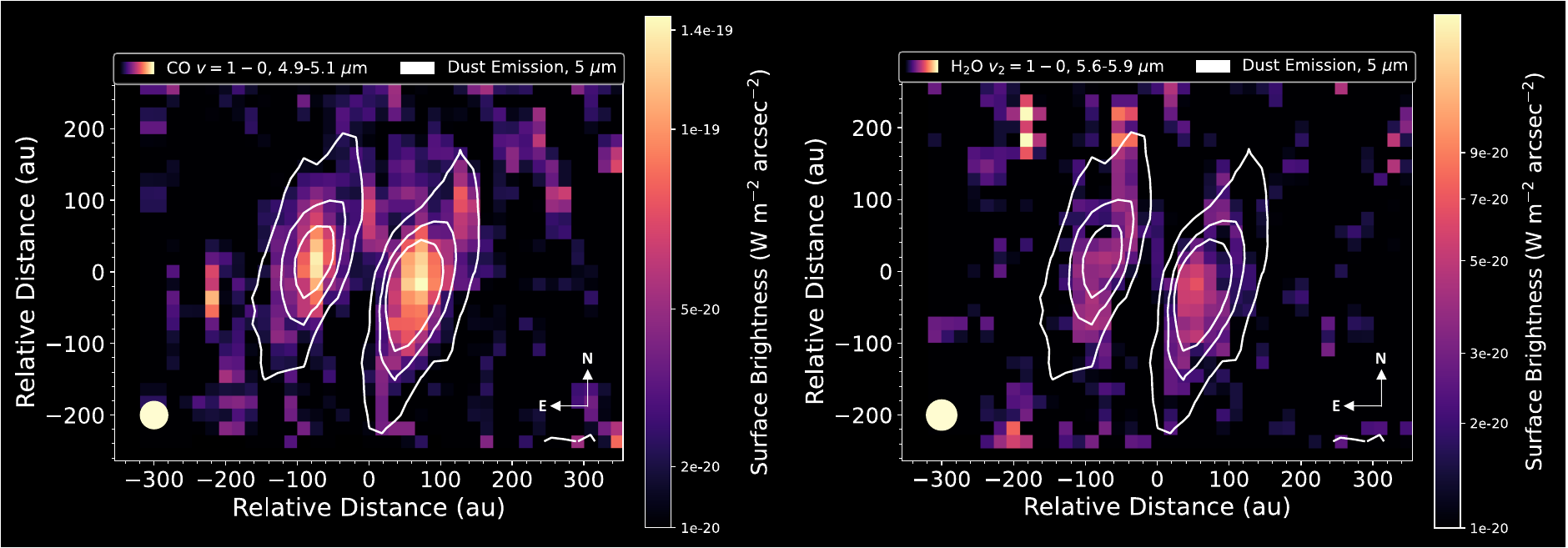}
\caption{MIRI MRS line images of Tau~042021, showing stacked emission lines from CO $v=1-0$ (left) and H$_2$O $v_2 = 1-0$ transitions, detected with the MIRI MRS. White contours in each image show the mid-IR scattered light continuum at 5.0 $\mu$m. Images have been scaled using an asinh stretch, to maximize the contrast between the background noise and the high S/N emission from the source. Beige circles in the bottom left corner of each panel represent the average theoretical FWHM of the PSF, calculated from the relationship reported in \citet{law2023}. The emission is spatially extended beyond the expected size of the inner disk $r < 1$ au, consistent with scattering by $\mu$m-sized dust grains in disk surface layers (see e.g., \citealt{Duchene24}).}
\label{fig:fig2_dustCOH2Obendmed}
\end{figure*}

\begin{figure*}
\centering
\includegraphics[width=\linewidth]{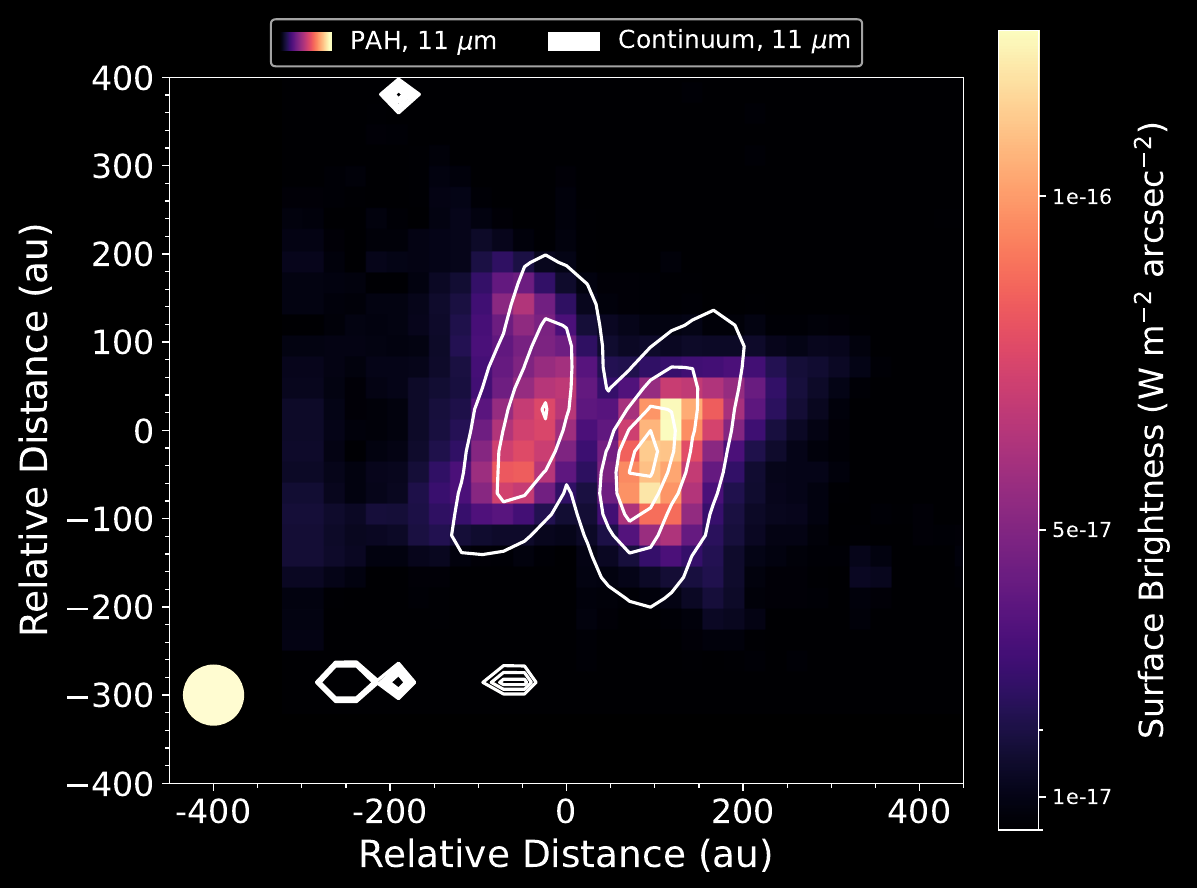}
\caption{MIRI MRS image of 11.3 $\mu$m PAH emission, which is spatially co-located with the scattered light continuum from Tau~042021. While PAH emission is readily detected in disks around more massive stars and in more externally irradiated environments (see e.g., \citealt{Geers2006, Geers2007}), it is rare to detect the band within disks around M-type stars. We note that the continuum on the red side of the band is beyond the edge of the Channel 2 detector, making it challenging to extract the total surface brightness profile. The image shown here is from a single frame where the feature peaks in emission.}
\label{fig:PAH}
\end{figure*}

In the MIRI MRS spectrum of Tau~042021, we detect the mid-infrared dust continuum and gas-phase emission lines from H$_2$ and [Ne II] that were first observed in the {\it Spitzer} spectrum (see Figure \ref{fig:1Dspec}, \citealt{Duchene24}). Additionally, we detect the 11.3 $\mu$m PAH band and gas-phase emission lines from CO (ro-vibrational), H$_2$O (rotational and ro-vibrational), OH (rotational), H I, [Ne III], [Fe II], [Ni II], [Ar II], and [S II]. The full list of detected transitions and emission line fluxes is provided in Table \ref{tab:line_IDs}. 

The 3-D spectral cube from Tau~042021 reveals spatially extended emission at wavelengths corresponding to the atomic and molecular gas-phase transitions detected in the 1-D spectra (see Figure \ref{fig:fig1_dustH2NeIIHI}, \ref{fig:fig2_dustCOH2Obendmed}). At the target distance of $d \sim 140$ pc, the pixel scales of the MIRI IFUs increase from $\sim$27 au (0.196'') in Channels 1 and 2 to $\sim$38 au (0.273'') in Channel 4 \citep{argyriou23}. Molecular gas in disks is not expected to produce strong mid-infrared emission lines at such large radial separations from the host stars (see e.g., \citealt{salyk2011}). However, recent MIRI MRS line images of FZ Tau revealed a spatially extended ring of H$_2$ emission \citep{Pontoppidan2023}. A ring of [Ar II] emission was also detected in the Beta Pictoris system, spatially coincident with its dusty debris disk \citep{worthen2024}. In the following sections, we present the continuum-subtracted line images of atomic and molecular lines in Tau~042021 and discuss the observed gas-phase morphologies. We find that the western nebula is brighter than the eastern nebula in all tracers reported in Table \ref{tab:line_IDs} and Table \ref{tab:spat_disk}, as expected if the western nebula is tilted slightly towards the observer \citep{Villenave2020, Simon19} or if the brightness of the eastern nebula is variable \citep{luhman2009}.

\subsection{Scattered and Thermal Dust Continuum Emission}
We observe a strong difference between the mid-infrared and millimeter emission of the disk. The scattered light lobes in Tau~042021 are similar to those seen with the \emph{I}-band filter on MegaPrime/MegaCam at the CFHT ($\lambda_0$ = 0.776 $\mu$m; \citealt{luhman2009}), the F606W ($\lambda_0 = 0.5888 \, \mu$m) and F814W ($\lambda_0 = 0.8115 \, \mu$m) filters of the Advanced Camera for Surveys (ACS) onboard the \emph{Hubble Space Telescope} \citep[][]{Stapelfeldt2014}, but require the presence of $10\,\mu$m grains at high disk altitudes \citep{Duchene24}. Conversely, the millimeter emission traces thermal emission from cold, large grains in the disk midplane. Each panel in Figure \ref{fig:fig1_dustH2NeIIHI} shows the 12 $\mu$m dust continuum from Tau~042021, drawn as white contours over the images. The upper left panel depicts the ALMA Band 7 (0.89 mm) emission from \citet{Villenave2020}. 

\subsection{Spatially Resolved Mid-IR H$_2$ Emission in a Protoplanetary Disk}

We detect 8 H$_2$ emission lines from the rotational ladder that fall within the MIRI MRS spectral range (see Table \ref{tab:line_IDs}). The upper right panel of Figure \ref{fig:fig1_dustH2NeIIHI} shows the image of the H$_2$ S(2) ($v = 0-0$) emission line at 12.278 $\mu$m, compared to the scattered light dust continuum at 12.2 $\mu$m. We observe H$_2$ emission across the entire radial extent of the dust disk, spanning a diameter of $220^{+8}_{-7}$ au in the western nebula. The H$_2$ S(2) emission is significantly more vertically extended than the dust scattered light, which only reaches $95^{+4}_{-3}$ au above the disk midplane. By contrast, the H$_2$ S(2) emerging from the western nebula extends to $171 \pm 7$ au. 

The vertically extended H$_2$ S(2) emission forms an X-shaped distribution, centered on the darker disk midplane. This is consistent with the \emph{JWST} NIRCam and MIRI images of Tau~042021, which revealed an X-shaped structure in the 7.7 and 12.8 $\mu$m filters that was attributed to a disk wind \citep{Duchene24}. We measured the semi-opening angle of each half of the X in the MIRI MRS line images, relative to the vertical disk axis (aperture position angle of 81.896$^{\circ}$), finding $\theta = 35^{\circ} \pm 5^{\circ}$. The angle with respect to the disk midplane is then $\theta \sim 55^{\circ}$, which is steeper than the $36^{\circ}$ angle between the midplane and the X in the 7.7 $\mu$m MIRI image \citep{Duchene24}. A similar ``onion-like" wind structure has previously been reported for DG Tau B, where velocity-resolved images show increasingly wide opening angles in emission lines tracing larger launching radii \citep{bacciotti2000, lavalley-fouquet2000, deValon2020, pascucci2023}. This includes the 45$^{\circ}$ semi-opening angle of the wide molecular wind traced by ro-vibrational H$_2$ $v = 1-0$ S(1) emission \citep{Beck2008, AgraAmboage2014} and the 90$^{\circ}$ cone measured from UV-fluorescent H$_2$ emission in the same system \citep{Schneider2013}. 

The 7.7 $\mu$m filter on MIRI spans wavelengths between 6.581-8.687 $\mu$m \citep{Wright2023}, which includes the H$_2$ S(5) and H$_2$ S(4) transitions detected in our MIRI MRS spectrum (6.909 $\mu$m, 8.025 $\mu$m, respectively). In the line images, the H$_2$ S(5), S(6), S(7), and S(8) emission lines trace an identical spatial region as the dust at 5 $\mu$m, while the H$_2$ S(1), S(3), and S(4) emission lines are also vertically extended. It is likely then that the X-shaped feature imaged at 7.7 $\mu$m by \citet{Duchene24} includes a superposition of the vertically extended H$_2$ S(4) and the less extended H$_2$ S(5) emission. We find that the sizes of both nebuluae decrease monotonically as a function of wavelength, with the H$_2$ S(7) emission line at 5.511 $\mu$m showing a height of just $94^{+5}_{-4}$ au and a diameter of $177 \pm 4$ au across the brighter, western lobe of the disk. This trend cannot be attributed to the increase in PSF size toward longer wavelengths, since the disk extends well beyond the the average PSF FWHM at the corresponding wavelengths listed in Table \ref{tab:line_IDs}. The spatially extended emitting regions we measure are also consistent with the 1-D spectroscopy results of \citet{Richter2002}, who reported non-detections of H$_2$ S(1)-S(8) rotational transitions within the small 2$\dprime$x2$\dprime$ aperture on the Texas Echelon Cross Echelle Spectrograph (TEXES) in a small sample of protoplanetary disks, despite clear detections with the larger aperture on the \emph{Infrared Space Observatory} that could better capture extended emission \citep{Thi2001}.

We note that the H$_2$ emission lines in which the extended emission is most prominent, S(1) and S(2) at 17.036 and 12.278 $\mu$m, respectively, are both observed at wavelengths longer than the limit below which the MIRI cruciform artifact appears ($\lambda < 10-12$ $\mu$m; \citealt{gaspar2021}). The feature is a result of lower quantum efficiency at shorter wavelengths in the MIRI detectors, which leaves the remaining photons to scatter internally. Since the extended H$_2$ emission is observed at longer wavelengths, where the quantum efficiency of the detectors is higher, the images are not contaminated by the artifact. 

\movetabledown=3mm
\begin{longrotatetable}
\begin{deluxetable*}{cccccc|ccccc}
\tablecaption{Detected Gas Phase Emission Lines \label{tab:line_IDs}}
\tablehead{\colhead{Species} & \colhead{Transition} & \colhead{Wavelength} & \colhead{$E_{\rm{upper}}$} & \colhead{$F_{\rm{int, west}}$} & \colhead{$F_{\rm{int, east}}$} & \colhead{PSF} & \colhead{$h_{\rm{west}}$} & \colhead{$d_{\rm{west}}$} & \colhead{$h_{\rm{east}}$} & \colhead{$d_{\rm{east}}$} \\
\colhead{} & \colhead{} & \colhead{($\mu$m)} & \colhead{(K)} & \colhead{(W m$^{-2}$)} & \colhead{(W m$^{-2}$)} & \colhead{(au)} & \colhead{(au)} & \colhead{(au)} & \colhead{(au)} & \colhead{(au)}}
\startdata
H$_2$\tablenotemark{a}, & S(8) & 5.053 & 8677.1 & $7.3 \times 10^{-20}$ & $3.3 \times 10^{-20}$ & 38 & $94 \pm 4$ & $138 \pm 5$ & $49^{+6}_{-30}$ & $84^{+20}_{-30}$ \\
 $v = 0-0$ & S(7) & 5.511 & 7196.7 & $2.6 \times 10^{-19}$ & $1.2 \times 10^{-19}$ & 40 & $94^{+5}_{-4}$ & $177 \pm 4$ & $36^{+19}_{-24}$ & $24^{+68}_{-16}$ \\
 & S(6) & 6.108 & 5829.8 & $1.3 \times 10^{-19}$ & $5.9 \times 10^{-20}$ & 43 & $97 \pm 4$ & $179 \pm 4$ & $46^{+26}_{-12}$ & $65^{+46}_{-29}$ \\
 & S(5) & 6.909 & 4586.1 & $7.1 \times 10^{-19}$ & $3.6 \times 10^{-19}$ & 47 & $113 \pm 4$ & $182 \pm 4$ & $82^{+8}_{-7}$ & $151^{+13}_{-20}$ \\
 & S(4) & 8.025 & 3474.5 & $3.5 \times 10^{-19}$ & $2.1 \times 10^{-19}$ & 52 & $151 \pm 7$ & $190 \pm 6$ & $125 \pm 6$ & $200^{+16}_{-15}$ \\
 & S(3) & 9.667 & 2503.7 & $1.0 \times 10^{-18}$ & $6.6 \times 10^{-19}$ & 60 & $156^{+8}_{-7}$ & $207 \pm 6$ & $142 \pm 9$ & $277^{+10}_{-8}$ \\
 & S(2) & 12.278 & 1681.6 & $4.6 \times 10^{-19}$ & $3.2 \times 10^{-19}$ & 72 & $171 \pm 7$ & $220^{+8}_{-7}$ & $150^{+13}_{-12}$ & $284 \pm 11$ \\
 & S(1) & 17.036 & 1015.1 & $9.3 \times 10^{-19}$ & $6.6 \times 10^{-19}$ & 94 & $190^{+22}_{-10}$ & $250 \pm 8$ & $127^{+16}_{-15}$ & $237^{+25}_{-41}$ \\
\hline
CO & $v = 1-0$, P(25)-P(40) & 4.909-5.092 & 4725.1-7338.1 & $\left(1.5-7.3 \right) \times 10^{-20}$ & $\left(1.3-3.3 \right) \times 10^{-20}$ & 38 & $94 \pm 3$ & $187 \pm 6$ & $73^{+5}_{-13}$ & $150^{+22}_{-20}$ \\
\hline
H$_2$O, & $11_{7 \, 5} \rightarrow 10_{4 \, 6}$ & 14.1780 & 3340.5 & $1.7 \times 10^{-20}$ & $1.3 \times 10^{-20}$ & 80 & $70^{+9}_{-10}$ & $113^{+13}_{-12}$ & $92^{+9}_{-7}$ & $142^{+10}_{-12}$ \\
 Rotational & $15_{6 \, 10} \rightarrow 14_{3 \, 11}$ & 14.2093 & 3443.2 & $3.4 \times 10^{-20}$ & $2.6 \times 10^{-20}$ & 80 & \nodata & \nodata & \nodata & \nodata \\
 & $14_{5 \, 10} \rightarrow 13_{2 \, 11}$ & 14.8951 & 4198.7 & $1.5 \times 10^{-20}$ & $1.4 \times 10^{-20}$ & 84 & \nodata & \nodata & \nodata & \nodata \\
 & $14_{6 \, 9} \rightarrow 13_{3 \, 10}$ & 14.9229 & 4438.4 & $1.6 \times 10^{-20}$ & $1.6 \times 10^{-20}$ & 84 & \nodata & \nodata & \nodata & \nodata \\
 & $13_{3 \, 10} \rightarrow 12_{2 \, 11}$ & 15.6246 & 3474.2 & $1.9 \times 10^{-20}$ & $1.8 \times 10^{-20}$ & 87 & $280 \pm 5$ & $255^{+8}_{-9}$ & \nodata & \nodata \\
 & $12_{6 \, 7} \rightarrow 11_{3 \, 8}$ & 16.1148 & 3501.7 & $2.5 \times 10^{-20}$ & $2.5 \times 10^{-20}$ & 89 & \nodata & \nodata & \nodata & \nodata \\
 & $12_{5 \, 8} \rightarrow 11_{2 \, 9}$ & 17.1025 & 3273.7 & $1.2 \times 10^{-20}$ & $1.5 \times 10^{-20}$ & 94 & \nodata & \nodata & \nodata & \nodata \\
 & $13_{4 \, 9} \rightarrow 12_{3 \, 10}$ & 17.5043 & 3645.6 & $1.5 \times 10^{-20}$ & $1.7 \times 10^{-20}$ & 96 & \nodata & \nodata & \nodata & \nodata \\
\hline
H$_2$O\tablenotemark{b}, & $7_{5 \, 3} \rightarrow 7_{4 \, 3}$ & 5.6564 & 9058.9 & $2.6 \times 10^{-20}$ & $2.6 \times 10^{-20}$ & 41 & $61 \pm 3$ & $90 \pm 6$ & $40^{+6}_{-11}$ & $33^{+14}_{-7}$ \\
Bend Mode & $8_{4 \, 4} \rightarrow 8_{3 \, 5}$ & 5.8801 & 9129.8 & $2.2 \times 10^{-20}$ & $1.5 \times 10^{-20}$ & 42 & \nodata & \nodata & \nodata & \nodata \\
$v_2=1-0$ & $6_{0 \, 6} \rightarrow 5_{1 \, 5}$ & 5.8964 & 8106.2 & $1.3 \times 10^{-20}$ & $9.7 \times 10^{-21}$ & 42 & \nodata & \nodata & \nodata & \nodata \\
 & $8_{1 \, 8} \rightarrow 8_{2 \, 7}$ & 6.9612 & 8661.5 & $2.5 \times 10^{-20}$ & $2.3 \times 10^{-20}$ & 47 & $<25$ & $<34$ & \nodata & \nodata \\
 & $3_{2 \, 1} \rightarrow 4_{3 \, 2}$ & 7.0228 & 7793.6 & $2.3 \times 10^{-20}$ & $9.6 \times 10^{-21}$ & 47 & \nodata & \nodata & \nodata & \nodata \\
 & $4_{3 \, 2} \rightarrow 5_{3 \, 3}$ & 7.1892 & 8051.2 & $1.1 \times 10^{-20}$ & $3.8 \times 10^{-21}$ & 48 & \nodata & \nodata & \nodata & \nodata \\ 
\hline
OH, & $v = 1-1$ R(18.5) & 14.650 & 10760 & $1.2 \times 10^{-20}$ & $8.2 \times 10^{-21}$ & 83 & $106 \pm 6$ & $148 \pm 7$ & \nodata & \nodata \\
Rotational & $v = 0-0$ R(20.5) & 13.54 & 7494.7 & $3.0 \times 10^{-21}$ & \nodata & \nodata & \nodata & \nodata & \nodata & \nodata \\
& $v = 1-1$ R(22.5) & 17.769 & 7082 & $1.8 \times 10^{-20}$ & \nodata & 97 & \nodata & \nodata & \nodata & \nodata\\
\hline
H I & Pf$\alpha$ (6-5) & 7.456 & 153420.9 & $6.5 \times 10^{-20}$ & $4.6 \times 10^{-20}$ & 50 & \nodata & \nodata & \nodata & \nodata \\
 & Hu$\beta$ (8-6) & 7.503 & 155338.6 & \nodata & \nodata & 50 & $12 \pm 7$ & $14 \pm 8$ & \nodata & \nodata \\
 & Hu$\alpha$ (7-6) & 12.372 & 154583.8 & $2.6 \times 10^{-20}$ & $1.9 \times 10^{-20}$ & 72 & $92 \pm 7$ & $113 \pm 13$ & $134 \pm 2$ & $121 \pm 3$ \\
\hline
[Ne II] & $^2P_{1/2}$--$^2P_{3/2}$ & 12.814 & 1122.9 & $1.3 \times 10^{-18}$ & $6.6 \times 10^{-19}$ & 74 & $145^{+8}_{-9}$ & $103^{+5}_{-3}$ & <110 & <66 \\ 
$\left[ \rm{Ne \, III} \right]$ & $^3P_{1}$--$^3P_{2}$ & 15.555 & 924.9 & $8.8 \times 10^{-20}$ & $8.2 \times 10^{-20}$ & 87 & $167^{+13}_{-10}$ & $120 \pm 4$ & $152^{+13}_{-12}$ & $114^{+20}_{-5}$ \\
\hline
[Fe II]\tablenotemark{c} & $^4F_{9/2}$--$^6D_{9/2}$ & 5.340 & 2694.2 & $1.3 \times 10^{-19}$ & $6.7 \times 10^{-20}$ & 40 & $150 \pm 15$ & $49 \pm 5$ & $150 \pm 15$ & $47 \pm 3$ \\
 & $^4F_{7/2}$--$^4F_{9/2}$ & 17.936 & 3496.4 & $6.2 \times 10^{-19}$ & $4.7 \times 10^{-19}$ & 98 & $<132$ & \nodata & $<96$ & \nodata \\
 & $^4F_{5/2}$--$^4F_{7/2}$ & 24.519 & 4083.2 & $2.1 \times 10^{-19}$ & $2.0 \times 10^{-19}$ & 128  & \nodata & \nodata & \nodata & \nodata \\
\hline
[Ni II] & $^2D_{3/2}$--$^2D_{5/2}$ & 6.636 & 2168.2 & $1.2 \times 10^{-19}$ & $4.5 \times 10^{-20}$ & 45 & $85 \pm 3$ & $68 \pm 5$ & \nodata & \nodata \\
 & $^4F_{7/2}$--$^4F_{9/2}$ & 10.682 & 13423.9 & $3.9 \times 10^{-20}$ & $2.2 \times 10^{-20}$ & 64 & $167 \pm 15$ & $<73$ & $107^{+13}_{-5}$ & $75^{+2}_{-3}$ \\
 & $^2F_{5/2}$--$^2F_{7/2}$ & 12.729 & 14554.3 & $1.1 \times 10^{-20}$ & $8.8 \times 10^{-21}$ & 74 & $44 \pm 15$ & $24 \pm 20$ & \nodata & \nodata \\
\hline
[Ar II] & $^2P^0_{1/2}$--$^2P^0_{3/2}$ & 6.985 & 2059.8 & $2.1 \times 10^{-19}$ & $8.5 \times 10^{-20}$ & 47 & $84^{+3}_{-4}$ & $72^{+3}_{-4}$ & \nodata & \nodata \\
\hline 
[S III] & $^3P_{2}$--$^3P_{1}$ & 18.713 & 1198.6 & $3.6 \times 10^{-20}$ & $4.4 \times 10^{-20}$ & 101 & \nodata & \nodata & \nodata & \nodata \\
\enddata
\tablenotetext{a}{Upper level energies (E$_{upper}$) for H$_2$ from \citet{Roueff2019}}
\tablenotemark{b}{Upper level energies (E$_{upper}$) for H$_2$O (rotational and bend mode) from \citet{Barber2006}}
\tablenotetext{c}{Upper level energies (E$_{upper}$) for [Fe II] from \citet{Bautista15}}
\tablenotetext{*}{For the forbidden emission lines, $h$ represents the vertical extent of the knots closest to the midplane and $d$ represents the jet widths.}
\end{deluxetable*}
\end{longrotatetable}

\begin{deluxetable*}{cc|cccc}
\tablecaption{Measured Spatial Extents of Dust Emission \label{tab:spat_disk}}
\tablehead{\colhead{Tracer} & \colhead{PSF FWHM} & \colhead{$h_{\rm{west}}$} & \colhead{$d_{\rm{west}}$} & \colhead{$h_{\rm{east}}$} & \colhead{$d_{\rm{east}}$} \\
 & \colhead{(au)} & \colhead{Height} & \colhead{Diameter} & \colhead{Height} & \colhead{Diameter}
}
\startdata
Dust, Channel 1\tablenotemark{a} & 38 & $95^{+4}_{-3}$ & $181^{+11}_{-10}$ & $67 \pm 7$ & $135 \pm 13$ \\
Dust, Channel 2\tablenotemark{b} & 60 & $139 \pm 6$ & $233^{+11}_{-10}$ & $121 \pm 8$ & $219^{+15}_{-20}$ \\
Dust, Channel 3\tablenotemark{*}\tablenotemark{c} & 94 & $302^{+10}_{-9}$ & $301^{+15}_{-16}$ & \nodata & \nodata \\
Dust, Channel 4\tablenotemark{*} & 98 & $321 \pm 10$ & $304^{+15}_{-14}$ & \nodata & \nodata \\
\enddata
\tablenotetext{*}{No resolved separation between eastern and western lobes}
\tablenotetext{a}{Channel 1: $\lambda = 4.9-7.65$ $\mu$m; Channel 2: $\lambda = 7.51-11.7$ $\mu$m; Channel 3: $\lambda = 11.55-17.98$ $\mu$m; Channel 4: $\lambda = 17.7-27.9$ $\mu$m}
\tablenotetext{b}{Medium Sub-Band: $\lambda = 5.66-6.63$ $\mu$m; Long Sub-Band: $\lambda = 6.53-7.65$ $\mu$m}
\tablenotetext{c}{Medium Sub-Band: $\lambda = 13.34-15.57$ $\mu$m; Long Sub-Band: $\lambda = 15.41-17.98$ $\mu$m}
\end{deluxetable*}

\subsection{Bipolar Jets Detected in Forbidden Emission}
We report the detection of spatially extended [Ne II] (12.814~$\mu$m) emission from Tau~042021 (see Figure \ref{fig:fig1_dustH2NeIIHI}). Emission lines from [Ne III] (15.555~$\mu$m), [Fe II] (5.34, 17.936~$\mu$m), [Ni II] (6.636, 10.682, 12.729, 18.241~$\mu$m), [Ar II] (6.985~$\mu$m), and [S III] (18.713~$\mu$m) are also identified and spatially resolved, with median emitting regions between $\sim67-75$~au wide and $\sim$150~au long (see Figure \ref{fig:1Dspec}, Table \ref{tab:line_IDs}). Like the H$_2$ S(2) emission shown in Figure \ref{fig:fig1_dustH2NeIIHI}, the [Ne II] emission extends beyond the vertical height of both dust reflection nebulae observed at 12 $\mu$m. However, the [Ne II] appears in 4 distinct knots, with a much narrower radial extent than the H$_2$ in the knots that coincide with the dust continuum. The distribution is consistent with emission from a bipolar jet (see e.g., \citealt{Dougados2000}), which was discovered in the earlier \emph{HST} and \emph{JWST} images of Tau~042021 \citep{Duchene14, Duchene24}. 

High-resolution spectroscopic studies of mid-IR [Ne II] emission typically detect either a high-velocity component (HVC) associated with collimated jets or a low-velocity component (LVC) tracing either MHD or photoevaporative disk winds in the 1-D emission line profiles \citep{Herczeg07, Gudel2010, Pascucci2011, Pascucci2020, Alexander2014} and 2-D line images \citep{bajaj2024}. Although the MIRI MRS observations do not have the spectral resolution to distinguish HVC or LVC components, the X-shape observed with the F1280W broadband filter on MIRI \citep{Duchene24} is consistent with the H$_2$ S(2) emission that we observe, and not with the [Ne II] jet. We do not spatially resolve changes in the jet width as a function of distance from the star, which places an upper limit of $<3^{\circ}$ on the jet opening angle (see e.g., \citealt{Narang2023}). Figure \ref{fig:fig1_dustH2NeIIHI} shows that the position angle of the [Ne II] jet appears slightly offset from the position angle of the sub-mm disk ($\rm{PA} = -16^{\circ}$; \citealt{Villenave2020}), and we measure a shift of $\sim 2.5^{\circ}$. We discuss possible sources for the misalignment in Section 4.2.

\subsection{Detection of spatially extended PAH emission}
The PAH emission feature seen in the {\it Spitzer} spectrum and detected in Figure 1 is also extended with approximately the same distribution as the scattered dust continuum. The IFU imaging proves that the PAH emission is absent in the background and localized to the disk. With a spectral type of M1$\pm$2 \citep{luhman2009}, this may be the first detection of PAH emission around a low mass T Tauri star (M-K spectral type) where the PAH emission is intrinsic to the disk itself rather than due to external irradiation \citep{Geers2006}. It may be that the unique geometry of this system, which blocks the thermal emission from the inner disk, allows the relatively weaker PAH emission around T Tauri stars to be detected at mid-infrared wavelengths. The spatial extension of the PAH feature is robust; however its location at the end of channel 2 means that the red continuum is poorly defined. Therefore we leave the analysis of this feature to future work.

\subsection{Scattering of Inner Disk CO and H$_2$O Emission}
Spatially resolved emission from CO ro-vibrational lines between $4.9-5.3$ $\mu$m is detected in the MIRI MRS cube, corresponding to the transitions $v = 1-0$ P(25)-P(40). Figure \ref{fig:fig2_dustCOH2Obendmed} shows the combined emission from these features after subtraction of the local continuum, revealing a span of $187 \pm 6$~au in diameter and $94 \pm 3$ au in height above the disk midplane for the western lobe of the disk. The fainter, eastern lobe has a diameter of $150^{+22}_{-20}$~au and a height of $73^{+5}_{-13}$~au. By contrast, spectro-astrometric observations of the ro-vibrational CO lines were consistent with more compact distributions of hot gas, with LTE temperatures of 100-1000~K and emitting areas inside $r < 4$~au \citep{Pontoppidan2011, Brown2013}. The emitting region we observe with MIRI in this edge-on disk is significantly larger than expected for lines originating in gas within the 100-1000~K temperature range, indicating that the extended emission is more likely scattered light (see e.g., \citealt{Pontoppidan2002}).

Numerous emission lines from the rotational and ro-vibrational bending mode transitions of H$_2$O (5.65-6.19 $\mu$m; see Table \ref{tab:line_IDs}) are also readily detected in the MIRI spectral cube. Figure \ref{fig:fig2_dustCOH2Obendmed} shows the stacked images of ro-vibrational bending mode features detected within the medium sub-band of Channel 1. The H$_2$O distributions extend far beyond the emitting regions derived from slab model fits to mid-infrared spectra from less inclined disks ($r<1$ au; see e.g., \citealt{salyk2011, Perotti23, banzatti2023, gasman2023, munozromero2024}). The image of the H$_2$O rotational lines in Tau~042021 shows a diameter of $113^{+13}_{-12}$ au and a vertical extent of $70^{+9}_{-10}$ au in the western lobe of the disk. In Channel 1, the ro-vibrational lines between 5.6-5.9 $\mu$m span a disk diameter of $90 \pm 6$ au in the bright lobe and a vertical height of $61 \pm 3$ au. However, the larger pixel scale in Channel 3 makes it challenging to directly compare the spatial distributions of the rotational and  bending mode emission lines, and the difference in emitting areas reported here is not statistically significant.

\subsection{Hydrogen Recombination Lines Emerging from Accretion Shocks}
The bottom right panel of Figure \ref{fig:fig1_dustH2NeIIHI} shows the 2-D spectral image of the H I (7-6) Hu$\alpha$ emission line at 12.372 $\mu$m. We also identify H I Pf$\alpha$ (7.456 $\mu$m) and Hu$\beta$ (7.503 $\mu$m) emission lines in the 1-D spectrum, which were tentatively detected in the \emph{Spitzer}-IRS spectrum \citep{Duchene24}. However, the peak surface brightnesses from the Pf$\alpha$ and Hu$\beta$ emission lines ($\sim 5 \times 10^{-20}$ W m$^{-2}$ arcsec$^{-2}$) are close to the background noise level in the continuum-subtracted images, making it challenging to measure the spatial extent of the emitting regions. We constrain the stronger H I (7-6) emission to a diameter of $113 \pm 13$ au and a height of $92 \pm 7$ au in the brighter, western nebula. The fainter, eastern nebula has a diameter of $121 \pm 3$ au and a height of $134 \pm 2$ au. 

Infrared H I recombination lines are typically correlated with accretion luminosity in Class I (see e.g., \citealt{Fiorellino2021}) and Class II sources (see e.g., \citealt{Pascucci07, Salyk2013, Rigliaco2015, Alcala2017}), implying that the emission originates within the accretion shocks rather than the disk surface. In particular, the Hu$\alpha$ transition was detected in 46/114 disks observed with \emph{Spitzer} (40\%), where it was found to trace H I number densities between $10^{10}-10^{11}$ cm$^{-3}$ \citep{Rigliaco2015}. We measure the integrated flux in the H I (7-6) emission line from Tau~042021, finding $F = 2.6 \times 10^{-20}$ W m$^{-2}$ in the western lobe and $1.9 \times 10^{-20}$ W m$^{-2}$ on the eastern side of the disk. However, Figure \ref{fig:fig1_dustH2NeIIHI} shows that much of the intrinsic emission line flux is likely scattered out of the line-of-sight. 

In a less inclined disk, the infrared SED is relatively flat between $\sim$12-100 $\mu$m (see e.g., \citealt{dullemond2007}). The SED from Tau~042021 shows that the flux increases from $\lambda F_{\lambda} \sim 10^{-16}$ W m$^{-2}$ at $\sim$12 $\mu$m to $\lambda F_{\lambda} \sim 1 \times 10^{-14}$ W m$^{-2}$ at $\sim$ 100 $\mu$m \citep{Duchene24}, as the contribution from thermal emission becomes visible in the edge-on disk. If the decrease is predominantly an inclination effect, the ``true" flux at $\sim$12 $\mu$m is likely also closer to $10^{-14}$ W m$^{-2}$. We include this flux ratio of $\sim$100 as a scaling factor to correct the H I Hu$\alpha$ emission line flux for scattering and use the linear relationship between the H I and accretion luminosities \citep{Rigliaco2015} to derive $L_{\rm{acc}} = 7.9 \times 10^{-5} \, L_{\odot}$ for Tau~042021. The accretion rate can then be estimated as $\dot{M}_{\rm{acc}} = L_{\rm{acc}} R_{\ast} / G M_{\ast} \left(1 - R_{\ast} / R_{\rm{in}} \right) = 2 \times 10^{-11} \, M_{\odot}$ yr$^{-1}$ \citep{Gullbring98}, assuming $R_{in} = 5 R_{\ast}$ \citep{Rigliaco2015}, $R_{\ast} = 1.7 \, R_{\odot}$, and $M_{\ast} = 0.27 \, M_{\odot}$ \citep{Simon19}. This value is likely a significant underprediction, and we discuss the implications further in Section 4.2. 

\section{Analysis}

\subsection{Extracting the Molecular Gas Thermal Structure}

To explore whether the radially extended H$_2$, CO, and H$_2$O emission is consistent with a distribution of scattered light, we used the spectools\textunderscore ir package \citep{salyk2022} to derive the temperature, column density, and emitting area of the warm gas. spectools\textunderscore ir produces slab models in Local Thermodynamic Equilibrium (LTE), which we fit to the 1-D spectrum of Tau~042021 after using the publicly available ctool routine \citep{Pontoppidan2023} to subtract the continuum. We fit each molecule individually, as the H$_2$ and CO emission lines are well separated from both the rotational and ro-vibrational H$_2$O transitions. 

Table \ref{tab:retrievals} reports the best-fit slab model parameters for all three molecules. We find that the CO spectrum is consistent with $T = 1500$ K and $\log N = 17.5$ cm$^{-2}$, demonstrating that the temperature required to produce the $J = 25-40$ emission lines is  similar to what is observed in less inclined disks \citep{salyk2011, Brown2013}. A similar effect is seen in the H$_2$O emission lines, which require $T = 850$ K and $\log N = 15.0$ cm$^{-2}$. Again, this is roughly consistent with \emph{JWST} MIRI MRS observations of water in less inclined protoplanetary disks, which show temperatures between $\sim 600-900$ K \citep{Kospal2023, Grant2023, gasman2023, Perotti23, banzatti2023, munozromero2024}. We note that Tau~042021 does not show an excess cool H$_2$O component, which has been associated with pebble drift in compact disks \citep{banzatti2023}. Rather than originating at large radial and vertical separations, the extended emission we observe is likely scattered along the disk surface by large dust grains, an effect uniquely detected in edge-on systems. 

In the case of H$_2$, we find that two temperature components are required to reproduce the observed spectrum. The first, with $T = 400$ K and $\log N = 19.4$ cm$^{-2}$ captures the bulk of the emission from the S(1)-S(4) emission lines, which are vertically extended in the line images. However, this model under-predicts the S(5)-S(8) emission line fluxes, which are better reproduced when an additional gas population with $T = 1500$ K and $\log N = 20.1$ cm$^{-2}$ is included. The hotter temperature is consistent with the CO within the inner disk, while the cooler temperature is closer to the 100 K wind temperature assumed for the mid-infrared H$_2$ emission lines measured by \citet{Thi2001}. However, we note that the emitting area we retrieve for the 1500 K H$_2$ ($r = 5$ au) is significantly larger than the intrinsic emitting area of the CO ($r = 0.01$ au).

\begin{deluxetable*}{cccc}
\tablecaption{Slab Model Retrieval Parameters \label{tab:retrievals}}
\tablehead{\colhead{Species} & \colhead{T (K)} & \colhead{$\log N$ (cm$^{-2}$)} & \colhead{r (au)}
}
\startdata
CO & 1500 & 17.5 & 0.01 \\
H$_2$O & 850 & 15.0 & 0.8 \\
H$_2$, hot & 1500 & 20.1 & 5 \\
H$_2$, warm & 400 & 19.4 & 175\tablenotemark{a} \\
\enddata
\tablenotemark{a}{We allowed the emitting area of the 400 K H$_2$ component to vary between 0.01-200 au, based on the vertically extended spatial distribution in Figure \ref{fig:fig1_dustH2NeIIHI}.}
\end{deluxetable*}

\subsection{No Velocity Structure Detected in Spatially Extended Emission Lines}

\begin{figure*}
\centering
\includegraphics[width=\linewidth]{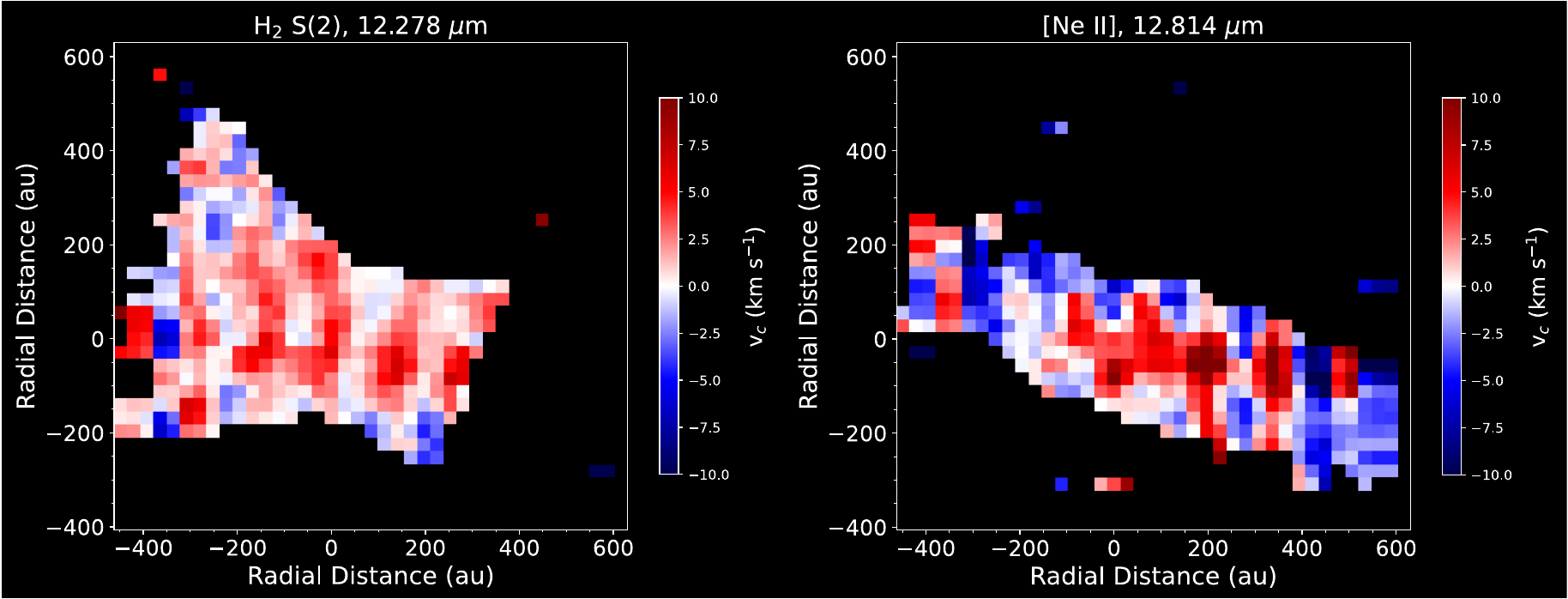}
\caption{Emission-weighted velocity maps of H$_2$ S(2) \emph{(left)} and  [Ne II] emission lines from Tau~042021 \emph{(right)} (see e.g., \citealt{Beck2008, Narang2023}). No velocity structure is detected in either of the maps, as expected for an edge-on disk from which the maximum jet and wind velocity components are not oriented along the line-of-sight.}
\label{fig:vc_maps}
\end{figure*}

We explore the velocity structure in the H$_2$ and atomic emission lines, using both the 1-D MIRI MRS spectrum and the spectral images. We fit each of the emission lines listed in Table \ref{tab:line_IDs} with a model of a Gaussian profile, superimposed on a linear continuum extending $\pm 0.02$ $\mu$m from the central wavelength of the transition. The free parameters are the amplitude, velocity centroid, and width of the Gaussian profile and the slope and intercept of the continuum. The best-fit velocity centroids for the H$_2$ emission lines range between $-20$ km s$^{-1}$ and $2$ km s$^{-1}$, with no trends as a function of wavelength. A similar range is derived from the fits to the atomic emission lines, which span $-25$ km s$^{-1}$ to 6 km s$^{-1}$. The best-fit FWHMs of the emission lines indicate that the features are barely wider than the instrument resolution, making it challenging to detect any kinematic signatures in the 1-D line profiles.

Figure \ref{fig:vc_maps} shows the emission-weighted velocity maps for the H$_2$ S(2) and [Ne II] emission lines (see e.g., \citealt{Beck2008, Narang2023}). No velocity gradients are detected in either of the spectral images. We do not identify a red- or blue-shifted side of the jet or wind, consistent with expectations for outflowing material moving perpendicularly away from the edge-on disk and therefore the line-of-sight. We also search for signatures of jet wiggling, which is readily detected over large spatial scales with VLT/MUSE and may be caused by perturbations from a misaligned companion or a precessing inner disk (see e.g., \citealt{Murphy2021}). We do not resolve any such velocity structure in the knots along the jet axis in Tau~042021, although we note that the difference in surface brightness between the eastern and western disk nebula has previously been reported as variable and may also be characteristic of a misaligned inner disk \citep{luhman2009}. Larger mosaics and multi-epoch observations will be required to fully characterize jet precession and emission variability in MIRI-MRS line images.

Observations of near-IR H$_2$, [O I] $\lambda$6300 \AA \, and high-resolution [Ne II] emission lines have blue-shifted wind components with velocity centroids $v_c < 20$ km s$^{-1}$ \citep{Pascucci2009, Pascucci2020, Gangi2020, Beck2019, Banzatti2019}, which is smaller than the spectral resolution of MIRI at 12 $\mu$m ($\Delta v \sim 120$ km s$^{-1}$). For a typical jet speed of $\sim150-200$ km s$^{-1}$ \citep{shang1998, Narang2023}, the line-of-sight velocity toward a disk with $i > 85^{\circ}$ \citep{Villenave2020} would be $<20$ km s$^{-1}$. This is again smaller than the MIRI spectral resolution and consistent with the velocity centroids in Figure \ref{fig:vc_maps}. Rather, the images of spatially extended wind and jet emission presented here are complementary to high spectral resolution observations that can measure kinematics. 

\subsection{Comparing Mass Loss Rates Derived from MIRI MRS Spectral Images}

As reported in Section 3, vertically extended H$_2$ is observed beyond the scattered light continuum in the S(1)-S(4) emission lines. This behavior is distinctly different from the CO and H$_2$O emission lines, which show emitting regions that are more consistent with the dust distributions in the MIRI MRS wavelength range. Assuming that the observed H$_2$ S(1) emission is optically thin, we use the integrated line flux to estimate the mass of warm gas contained in the wind \citep{Thi2001}:
\begin{equation}
F_{J'-J''} = \frac{h c}{4 \pi \lambda} N \left(\rm{H}_2 \right) A_{J'-J''} x_{J'} \Omega,
\end{equation}
where $x_{J'} = g_u \times \exp{(-E_{J'-J''} / k T_{\rm{exc}})} / Q(T_{\rm{exc}})$ is the fraction of H$_2$ molecules in the upper state $\left(J = 3 \right)$, $Q(T_{\rm{exc}})$ is the partition function of H$_2$ at temperature $T_{\rm{exc}}$ \citep{Popovas2016}, $A_{J'-J''}$ is the Einstein coefficient describing the probability of radiative decay from $J = 3$ to $J = 1$ \citep{Roueff2019}, and $\Omega$ is the solid angle spanned by the emitting region reported in Table \ref{tab:spat_disk}. 

The H$_2$ emission lines in the MIRI MRS spectrum are consistent with two temperature components: $T = 400$ K, which dominates the S(1)-S(4) emission lines, and $T = 1500$ K to capture the line fluxes from S(5)-S(8). Assuming the 400 K temperature component in the wind, we measure $M_{{\rm H}_2} \sim 5.4 \times 10^{-7} \, M_{\odot}$. At the typical wind velocities of $\sim$10 km s$^{-1}$ predicted by models and the measured maximum emission height of 175 au above the disk midplane, this corresponds to a mass flux of $\dot{M}_{\rm{wind}} \sim 6.5 \times 10^{-9} \, M_{\odot}$ yr$^{-1}$. The uncertainties in this measurement are dominated by the size of the emitting region and the gas temperature, which is less constrained without observations of the H$_2$ S(0) emission line at 28.2 $\mu$m. 

The measured accretion rate, $\dot{M}_{\rm{acc}} = 2 \times 10^{-11} \, M_{\odot}$ yr$^{-1}$ is approximately two orders of magnitude lower than the wind mass loss rate, which contradicts the ratios of $\dot{M}_{\rm{wind}} / \dot{M}_{\rm{acc}} \sim 0.1-1$ measured from the $^{12}$CO and [O I] $\lambda$6300 LVC emission lines \citep{pascucci2023}. However, we note that the measured accretion rate is also three orders of magnitude lower than the threshold for strong accretors with HVC [Ne II] emission line components in 1-D spectra \citep{Pascucci2020}, which originate in jets like the one imaged with MIRI from Tau~042021. This apparent contradiction likely takes its origin in the fact the the H I line may be strongly impacted by scattering along the line of sight through the edge-on disk, masking the bulk of the emission from the accretion shock itself (see e.g., \citealt{Alcala2019}). Additionally, some of the Hu$\alpha$ emission may originate at the base of the jet, and the large methodological uncertainty in deriving a stellar mass for a target in an edge-on disk plays a role as well (see e.g., \citealt{Hartmann16} for a recent review). However, we note that the line-to-continuum ratios across both sides of the disk are roughly constant, with $L/C \sim 1.3 \times 10^{-4}$ in the eastern nebula and $L/C \sim 2.8 \times 10^{-4}$ in the western lobe. This effect is not seen in the atomic forbidden line images tracing the more extended jet. It is likely that the linear relationship between H I Hu$\alpha$ emission and accretion luminosity does not hold for edge-on disks like Tau~042021, which will require H I and dust scattering models to retrieve the missing line flux.     

\subsection{Distinguishing Between Outflow-Launching Mechanisms}

The semi-opening angle of $35^{\circ} \pm 5^{\circ}$ measured from the H$_2$ S(2) emission line (Figure \ref{fig:fig1_dustH2NeIIHI}) is consistent with the range of semi-opening angles between $\sim20-40^{\circ}$ for low-velocity winds reported in \citet{pascucci2023}. These values were derived from spectrally resolved [O I] $\lambda$ 6300 emission lines, which can be decomposed into high-velocity ($v = 50-300$ km s$^{-1}$ and low-velocity components ($v < 10$ km s$^{-1}$). The largest blueshifts are detected in low-velocity components from disks with inclinations $i \sim 35^{\circ}$, where the ejected material is moving directly toward the observer along the line-of-sight \citep{Banzatti2019}. Indeed, MHD wind models predict that mass outflow occurs when the poloidal magnetic field lines reach semi-opening angles of $\theta < 30^{\circ}$ relative to the disk rotation axis \citep{Blandford1982}. Our H$_2$ observations are consistent with both the wind model predictions and the [O I] low-velocity components, implying a MHD wind origin for the vertically extended emission.

When spatially resolved, the launching radii of the H$_2$ wind and the atomic jet can provide further constraints on the outflow-launching mechanism. A compact emitting region is typically associated with an MHD wind ($r < 1$ au; \citealt{Fang2023}), while more extended distributions are characteristic of photoevaporative winds \citep{bajaj2024} and jets \citep{Whelan2021, Murphy2021}). However, the pixel scale of the MIRI Channel 3 detector (0.245'', or 34 au at $d = 140$ pc) is large relative to the expected launching radii, and we do not meaningfully distinguish anything inside this minimum inner working angle.

\emph{Spitzer}-IRS observations of Tau~042021 were used to classify the object as a Class II disk with no remaining envelope, based on its extinction-free infrared spectral indices ($n_{5-12} = -0.73$, $n_{12-20} = -0.75$; \citealt{mcclure2010, furlan2011}). Although $^{13}$CO $J=3-2$ emission is observed at large scale heights \citep{Simon19}, it is not coincident with the vertically extended emission near 12 $\mu$m \citep{Duchene24} and the sub-mm gas surface density distribution does not show signs of an outflow cavity. The mid-infrared H$_2$ emission also does not appear similar to the spatially resolved outflow cavities detected in younger protostellar systems (see e.g., \citealt{federman2023}), consistent with the classification of Tau~042021 to a later stage of disk evolution. As reported in Section 4.1, we do not resolve velocity structure in this edge-on disk with the MIRI MRS, but the measured semi-opening angle of the spatially extended H$_2$ emission is most consistent with MHD wind model predictions. This is consistent with models that require MHD winds to reproduce the observed relationship between $\dot{M}_{\rm{acc}}$ and $M_{\rm{disk}}$ in the disk population \citep{tabone2022, somigliana23}. 

\section{Summary \& Conclusions}

We present \emph{JWST} MIRI MRS observations of gas-phase emission in the edge-on protoplanetary disk system Tau~042021. The data provide spatially resolved spectral images of gas-phase atomic and molecular emission lines, revealing:

\begin{itemize}

\item An X-shaped MHD disk wind with a semi-opening angle of $35^{\circ} \pm 5^{\circ}$, traced by vertically extended emission from rotational transitions of H$_2$ (see also, \citealt{Duchene24});

\item A collimated, bipolar jet in emission from [Ne II], [Ne III], [Ni II], [Fe II], [Ar II], and [S III] (see e.g., \citealt{Duchene14, Duchene24}), with an opening angle of $<3^{\circ}$;

\item Radially extended CO and H$_2$O consistent with line emission from the inner disk at 1 au, scattered on large dust grains in the disk surface out to 100 au.

\item H I Hu$\alpha$ emission in each nebula, consistent with scattered light from the accretion shocks or emission from the base of the jet.

\end{itemize}

Assuming the H$_2$ emission is optically thin, the wind contains a gas mass of $M_{H_2} \sim 5.4 \times 10^{-7} \, M_{\odot}$, corresponding to a mass flux of $\sim 6.5 \times 10^{-9} \, M_{\odot}$ yr$^{-1}$. The Hu$\alpha$ emission was used to estimate a mass accretion rate of $\dot{M}_{\rm{acc}} = 2 \times 10^{-11} \, M_{\odot}$ yr$^{-1}$, although we note that this is likely an underestimate due to scattering and extinction along the line-of-sight and uncertainties in the size of the line emitting region. This work demonstrates the unexpected ability to spatially resolve gas-phase emission lines from large protoplanetary disks with the MIRI MRS, which we have shown to be a powerful tool for measuring disk mass loss in winds and jets and for exploring inner disk sub-structure and compositions through scattered light.     

\section{Acknowledgements}

This work is based on observations made with the NASA/ESA/CSA James Webb Space Telescope. The data presented in this article were obtained from the Mikulski Archive for Space Telescopes (MAST) at the Space Telescope Science Institute, which is operated by the Association of Universities for Research in Astronomy, Inc., under NASA contract NAS 5-03127 for JWST. These observations are associated with program \#1751. The specific observations analyzed can be accessed via \dataset[10.17909/mx4k-mv71]{https://doi.org/DOI}. Support for program \#1751 was provided by NASA through a grant from the Space Telescope Science Institute, which is operated by the Association of Universities for Research in Astronomy, Inc., under NASA contract NAS 5-03127. A portion of this research was carried out at the Jet Propulsion Laboratory, California Institute of Technology, under a contract with the National Aeronautics and Space Administration (80NM0018D0004). E.D. and J.A.N. acknowledge support from French Programme National `Physique et Chimie du Milieu Interstellaire' (PCMI) of the CNRS/INSU with the INC/INP, co-funded by the CEA and the CNES. M.N.D. acknowledges the Holcim Foundation Stipend, the Swiss National Science Foundation (SNSF) Ambizione grant number 180079, the Center for Space and Habitability (CSH) Fellowship, and the IAU Gruber Foundation Fellowship.
\bibliography{midas_gas}{}
\bibliographystyle{aasjournal}

\end{document}